\documentclass[12pt]{article}
\usepackage{epsfig,amssymb}
\usepackage{latexsym}

\newcommand{\bq}{\begin{equation}}
\newcommand{\eq}{\end{equation}}
\newcommand{\bqa}{\begin{eqnarray}}
\newcommand{\eqa}{\end{eqnarray}}
\newcommand{\ben}{\begin{enumerate}}
\newcommand{\een}{\end{enumerate}}
\newcommand{\bc}{\begin{center}}
\newcommand{\ec}{\end{center}}
\newcommand{\bqb}{\begin{eqnarray*}}
\newcommand{\eqb}{\end{eqnarray*}}

\def\pr#1#2#3{Phys. Rev. ${\bf{#1}}$, #2 (#3)}

\def\pl#1#2#3{Phys. Lett. ${\bf{#1}}$, #2 (#3)}

\def\np#1#2#3{Nucl. Phys. ${\bf{#1}}$, #2 (#3)}

\def\zp#1#2#3{Z. f. Phys. ${\bf{#1}}$, #2 (#3)}
\def\jhep#1#2#3{JHEP ${\bf{#1}}$, #2 (#3)}
\def\epj#1#2#3{Eur. Phys. J. ${\bf{#1}}$, #2 (#3)}
\def\ijmp#1#2#3{Int. J. Mod. Phys. ${\bf{#1}}$, #2 (#3)}

\def\jmp#1#2#3{J. Mod. Phys. ${\bf{#1}}$, #2 (#3)}

\begin{document}
\pagenumbering{arabic}
\thispagestyle{empty}
\def\thefootnote{\fnsymbol{footnote}}
\setcounter{footnote}{1}

\begin{flushright}
January 13, 2017\\
arXiv: 1611.02426\\
 \end{flushright}

\vspace{2cm}

\begin{center}
{\Large {\bf Remarkable signals of $t_R$ compositeness}}\\
 \vspace{1cm}
{\large G.J. Gounaris$^a$ and F.M. Renard$^b$}\\
\vspace{0.2cm}
$^a$Department of Theoretical Physics, Aristotle
University of Thessaloniki,\\
Gr-54124, Thessaloniki, Greece.\\
\vspace{0.2cm}
$^b$Laboratoire Univers et Particules de Montpellier,
UMR 5299\\
Universit\'{e} Montpellier II, Place Eug\`{e}ne Bataillon CC072\\
 F-34095 Montpellier Cedex 5, France.\\
\end{center}

\vspace*{1.cm}
\begin{center}
{\bf Abstract}
\end{center}

We are looking for remarkable differences between predictions of the
standard model and those of $t_R$ compositeness (while the $t_L$ remains elementary)
in various basic $t\bar t$ production processes,
without having to make difficult final polarization analyses.
We assume the presence of $q^2$ dependent form factors which first suppress the $t_R$
contributions at high $q^2$, and in addition, may also lead to an effective
$q^2$ dependent top mass. We further assume that these effects disappear
at low $q^2$, so that there are no anomalous couplings.
We show that large specific differences indeed appear in the high
energy limits of cross sections and asymmetries in $e^-e^+$, gluon+gluon,
$\gamma\gamma$ collisions and  other fusion processes and that this should
lead to a strategy for analyzing them.

\vspace{0.5cm}

\def\thefootnote{\arabic{footnote}}
\setcounter{footnote}{0}
\clearpage

\section{INTRODUCTION}

Motivated by several deficiencies of  SM, the possibility of compositeness
of particles previously considered as elementary has been considered since
a long time \cite{comp}. This idea has been applied to the Higgs sector and it has
also been assumed that a new world directly related to the Higgs boson and
connected to the ordinary fermions could be at the origin of the fermion masses.
The top quark, with its heavy mass,
would be especially concerned by this feature. The broad spectrum of fermionic
masses has suggested the possibility of partial compositeness where the size of
the mixing of an elementary fermion with a corresponding new state is related
to the value of its mass. The high mass of the top quark would correspond
to a large mixing value and even to the full $t_R$ compositeness. The $t_L$
state would remain elementary in order to not perturb the (pure left) $Wtb$
system. Models based on these assumptions have already appeared, see for example
\cite{Hcomp2,Hcomp3,partialcomp,Hcomp4}.

We do not intend to analyze such models in details but we want to suggest
basic and quick tests of the appealing idea that $t_R$ would be composite,
whereas $t_L$ would remain elementary.
We then show in which processes and through which observables these
basic tests could be performed, without making a difficult final polarization
analysis. Indeed, we will find places where spectacular
differences should appear between the usual elementary top case and the
above composite $t_{R}$, elementary $t_{L}$, case.

Our basic point is that a composite $t_R$ should have a form factor $F(q^2)$
such that its production through photon, Z, gluon point-like couplings should
be suppressed at high $q^2$. On the opposite, the elementary $t_L$ will
keep the usual Born coupling (with obviously higher order, small corrections that
we do not consider at this level of the study). The pure left $W$ couplings
will not be modified. The Higgs couplings (connecting $t_{R}$ and $t_{L}$)
are particularly model dependent and several types of modifications may appear.
Among them we will introduce the concept of effective top mass in processes where
the value of the top mass controls the behaviour of the amplitudes at high energy,
for example in those  involving Higgs boson or longitudinal $W,Z$.

We  apply these considerations to several $t\bar t$ production processes
in $e^-e^+$, gluon-gluon, $q\bar q$, $\gamma\gamma$, $WW$, $ZZ$
collisions, as well as in single top production.
The high energy limits of their observables
are computed in different options and indeed in some cases they
lead to spectacular differences (simple factors) between the SM and the
composite $t_{R}$ case, which could be observable
without making difficult final polarization measurements.

We insist on the basic aspect of our note: we are not discussing the existence
of anomalous couplings (their possible measurements at LHC and ILC are for example
discussed in \cite{Richard, Fabbrichesi}), but the presence of $q^2$ dependent form factors,
which do not modify the SM couplings at low $q^2$;
they would modify the $t_{R}$ production at high $q^2$
but not the $t_L$ one.
In the following,  we make illustrations using arbitrary test-form factors, just in order
to show how these high energy limits may be reached.
More detailed special studies and experimental tests could be done in the spirit
of \cite{Hff,Htt}.

The organization of the paper is the following. The $e^-e^+ \to t \bar t$
process is treated in Sec. II, $gg \to t \bar t$ and other hadronic processes
in Sec. III, photon-photon collision, WW and ZZ fusion and other processes in Sec. IV.
Conclusions and possible developments are given in Sec. V.\\

\section{$e^-e^+ \to \gamma,Z \to t \bar t$}

We modify the usual SM point-like $Vtt$ ($V=\gamma, Z$)  as
\bq
\bar u_t \gamma^{\mu}[g^L_{Vt}P_L+g^R_{Vt}P_R ]v_t
~~\to ~~\bar u_t \gamma^{\mu}[g^L_{Vt}P_L+g^R_{Vt}P_RF(s)] v_t~~, \label{eett-model}
\eq
\noindent
where  $P_{L/R}=(1\mp \gamma^5)/2$, the $t_L$ is kept elementary with its SM point-like couplings,
while  the $t_R$ compositeness introduces a form factor through the replacement
\bq
g^R_{Vt} \to g^R_{Vt} F(s) ~~ , \label{gRVt-s-replacement}
\eq
which for the consistency of the theory it is assumed to hold not only for  $V=\gamma, Z$, but
for the gluon also.
In the numerical illustration we  use the "test-form factor"
\bq
F(s)={4m^2_t+M^2\over s+M^2} ~~, \label{Fs-form}
\eq
where $M$ is a new physics scale fixed at $0.5$ TeV in the illustration.
Thus, $F(s)$ is equal to 1 at threshold and tends to 0 at high energy.
Our aim is to emphasize the high energy  properties when the $t_R$ contribution is suppressed.
No special meaning is  given to the precise form factor expression.
A global change of scale could be easily done if one wants to discuss the
observability at a very high energy collider.

The various observables (polarized, unpolarized cross sections and
asymmetries) are then computed with the usual formulas (see for example
\cite{bookee}). The observability of anomalous electroweak couplings in such
processes has been recently discussed in \cite{Richard,Barducci}.
But we now consider the ratios of the values of the observables in the composite $t_R$
case over the ones in the standard case. We want to emphasize the leading
effects of $t_R$ compositeness,
from the fact that it may imply (contrarily to the $t_L$ contribution)
that its corresponding photon and $Z$ couplings are suppressed at high energy, by
progressively vanishing above the new physics scale.
At this stage of the study, we ignore the 1 loop and higher order QCD or electroweak corrections.
These corrections affect slightly
the absolute values, but should not sensibly modify the ratios.

As one can see in Figure 1 , at high energy, (say above 1 TeV for $M=0.5$TeV)
we obtain the following values for the above mentioned ratios:
\begin{itemize}
\item
0.61 for the total unpolarized cross section ratios $ \sigma_{unp}$,

\item
0.88 for the longitudinally polarized cross section ratio $\sigma_{long}$,

\item
0.67 for the transverse polarization azimuthal factor $\sigma_{trans}$,

\item
1.17 for the unpolarized forward-backward asymmetry $A^{FB}_{unp}$,

\item
a longitudinally polarized forward-backward asymmetry $A^{FB}_{long}$, varying in the range
(1.25-2.25),

\item
and 2.81 for the longitudinal polarization asymmetry $A^{pol}_{long}$.

\end{itemize}

One can see that the most spectacular effects are already present
in the total unpolarized cross section,  but also in the
longitudinal $e^-e^+$ polarization asymmetry, as well as
in the azimuthal distribution in the case of transverse $e^-e^+$
polarization and the unpolarized forward-backward asymmetry.

Similar effects should be observed in
$\mu^+\mu^- \to t \bar t$. This process only differs by the presence
of the Higgs exchange in the s-channel but this contribution
is too weak to modify substantially the above results.\\

\section{EFFECTS IN HADRONIC COLLISIONS}

The main $t \bar t$ production mechanism is now $gg \to t \bar t$.
At tree level  there are 3 diagrams: (a) s-channel gluon exchange,
(b) top exchange in the t-channel,
(c) top exchange in the u-channel.
For recent analyses of this process with high order corrections see e.g. \cite{Kidonakis}
and for the search of anomalous couplings see \cite{Fabbrichesi}.
In our framework, we modify the $t_R$ couplings with the form factor $F(s)$ in (a),
$\tilde F(t)$ in (b) and  $\tilde F(u)$ in (c). Thus the total Born amplitude is written as
\bq
A^{Born}=A^{Born~a}+A^{Born~b}+A^{Born~c} ~~, \label{ggtt-model}
\eq
with the three terms respectively given by
\bqa
A^{Born~a} &= &- if^{ijl}{\lambda^l\over2}  {4\pi \alpha_s\over s}(\epsilon.\epsilon')
\Big [\bar u_t  \gamma^{\mu}(k-k')_{\mu}
\Big (P_L+F(s)P_R \Big )v_t \Big ] ~~, \label{ggtt-a-term} \\
A^{Born~b}&=&-~{4\pi \alpha_s\over
t-m^2_t}~ {\lambda^i\over2}{\lambda^j\over2}~
\bar u_t \Big [ \gamma^{\mu}\epsilon_{\mu}
 \gamma^{\nu}(p-k)_{\nu} \gamma^{\rho}\epsilon'_{\rho}
\Big (P_L+\tilde F^2(t)P_R \Big )\nonumber\\
&&+m_t \gamma^{\mu}\epsilon_{\mu} \gamma^{\rho}\epsilon'_{\rho} \tilde F(t) \Big ] v_t
~~, \label{ggtt-b-term} \\
A^{Born~c}&=&-~{4\pi \alpha_s\over
u-m^2_t}~ {\lambda^j\over2}{\lambda^i\over2}~
\bar u_t \Big [\gamma^{\mu}\epsilon'_{\mu}
 \gamma^{\nu}(p-k')_{\nu} \gamma^{\rho}\epsilon_{\rho}
\Big (P_L+\tilde F^2(u)P_R \Big )   \nonumber\\
&&+m_t \gamma^{\mu}\epsilon'_{\mu} \gamma^{\rho}\epsilon_{\rho} \tilde F(u) \Big ]
v_t  ~~, \label{ggtt-c-term}
\eqa
where the initial gluons have color indices $(i,j)$, momenta $(k,k')$,
polarization vectors $(\epsilon,\epsilon')$, while $p$ denotes the momentum
of the final $t$-quark and $P_{L/R}$ have already been defined immediately after
(\ref{eett-model}).

Precise prediction for the energy behaviour of this cross section will depend
on the choice of $t_R$ form factors $F(s)$ and $\tilde F(t)$, $\tilde F(u)$, where
one should use (\ref{Fs-form}) for $F(s)$,  while for a virtual top in the $t$ or $u$ channel
we should use
\bqa
g^R_{Vt}   & \to & g^R_{Vt} \tilde F(x) ~~, \label{gRVt-x-replacement} \\
\tilde F(x) &= & {M^2\over -x+M^2}~~, ~~x=t,u  ~~. \label{Fx-form}
\eqa
Doing so, we obtain in the central region the results shown in Figure 2, where  we give
the ratio of the modified
cross section over the standard one.
The curve is drawn for $\theta={\pi\over2}$ but it is valid for any other
central value, as soon as the corresponding $t,u$ values are comparable to  $s$.

This ratio should not be sensibly affected by QCD corrections.
As already mentioned such an arbitrary choice of form factor
has no special meaning,
we just want to insist on the fact that, at high energy, the
$t_R$ contribution which was equal to the $t_L$ one in SM
(apart from small differences due to electroweak corrections)
is suppressed by the form factors.
As one can see in Figure 2 this suppression leads indeed quickly to a
reduction of the cross section by a factor $1/2$.\\

\section{FUSION AND OTHER PROCESSES}

\subsection{$\gamma\gamma \to t \bar t$.}

Here we assume that real photon-photon collisions will be observable
at future colliders; for recent studies see  \cite{Badelek}.
The basic process is similar to gluon-gluon process $gg\to t \bar t$ considered
in Sec. III,  except for the absence of the s-channel term. In analogy to
(\ref{ggtt-b-term}, \ref{ggtt-c-term}), the $t-$ and $u-$ channel terms are respectively given
by
\bqa
A^{Born~b}&=&-~{e^2Q^2_t\over t-m^2_t}
\bar u_t \Big [\gamma^{\mu}\epsilon_{\mu}
 \gamma^{\nu}(p-k)_{\nu} \gamma^{\rho}\epsilon'_{\rho}
\Big (P_L+\tilde F^2(t)P_R \Big )\nonumber\\
&&+m_t \gamma^{\mu}\epsilon_{\mu}  \gamma^{\rho}\epsilon'_{\rho} \tilde F(t)\Big ]
v_t ~~, \label{gamggamtt-b-term} \\
A^{Born~c}&=&-~{e^2Q^2_t\over u-m^2_t}
\bar u_t \Big [ \gamma^{\mu}\epsilon'_{\mu}
 \gamma^{\nu}(p-k')_{\nu} \gamma^{\rho}\epsilon_{\rho}
\Big (P_L+\tilde F^2(u)P_R \Big )\nonumber \\
&&+m_t \gamma^{\mu}\epsilon'_{\mu} \gamma^{\rho}\epsilon_{\rho} \tilde F(u) \Big ]
v_t ~~ \label{gamggamtt-c-term}
\eqa
where we use the same $t_R$ form factors as above. The result for the ratio of
differential cross sections in the central region (for large $t,u$) is also illustrated
in Figure 2. As expected, we obtain also a reduction by a factor $1/2$ at high energy,
similarly to the $gg \to t \bar t$ case.

\subsection{FUSION PROCESSES in $e^+e^-$ OR HADRONIOC COLLISIONS}

We consider the $WW, ZZ , \gamma\gamma, \gamma Z \to t \bar t$ processes, for which
basic studies can be found in \cite{fusion1,fusion2,fusion3,fusion4}. Specifically, we  find:

\begin{itemize}

\item
(a) For $WW\to t \bar t$.

In this process the $t$- and $u$-channel bottom exchanges with the left-handed
W couplings have no form factor effect.
The $s$-channel  $\gamma,Z$ exchange contributions though, could be sensitive to the $t_R$ form factor,
but at high energy the right-handed part of the $\gamma+Z$ contribution cancels
(it corresponds to the pure left $W^3$ exchange). So finally the modifications will only  arise
from  the $s$-channel Higgs exchange connecting $t_L$ to $t_R$ and being proportional
to the top mass.
This contribution is totally model dependent, and we will consider two
extreme situations.

First, if this Higgs part retains its SM structure, the global $WW\to t \bar t$
amplitudes, dominated by the $t_L$ parts, will be very similar to the standard case.
In this case the ratio of cross sections to SM will be close to 1.

On the opposite, if we suppose that both the Higgs boson and the $t_R$ are composite,
whereas the $t_L$ remains elementary and point-like, then one can expect that at high
energy the $Ht_Lt_R$ coupling will be suppressed by the same type of form factor effects
which suppress the $\gamma t_Rt_R$ and $Zt_Rt_R$ couplings.

But this cannot be the only feature. The suppression of the $WW \to H\to t\bar t$
contribution will create a problem with the necessary cancellation of the
increasing $W_LW_L$ amplitudes
at high energy. Indeed at high energy in SM, the sum of the contributions due to $\gamma,Z$
exchange in the $s$-channel and bottom exchange in the $t$ and $u$ channel
is proportional to the top mass
and is canceled by the $H$ exchange in the $s$ channel. The suppression
of this last contribution should be accompanied by a suppression of the first sum.
The simplest solution to this problem could be to replace the fixed value of the
top mass by an effective mass (a kind of scale dependent mass in a way similar
to the QCD case, but much more violent, being due to compositeness),
i.e. in this picture we would replace the top mass as
\bq
m_t \to   m_t(s)=m_tF(s) ~~. \label{mtop-replacement}
\eq
This mass suppression affects both single and double longitudinal $W_L$
contributions (that are proportional to the top mass) and together with
the suppression of the Higgs exchange contribution leads to a strong reduction
of the cross section, as one can see by comparing   the left  and right panels  of Figure 3.
Note that in the left panel only the $g^R_{Vt}$ replacement effects (\ref{gRVt-s-replacement},
\ref{gRVt-x-replacement}) are used, while in the right panel the $m_t$ replacement
(\ref{mtop-replacement})
 is also included. As shown in the right panel of Figure 3,
 at high energy the ratio becomes of the order of 0.2.

\item
(b) $ZZ \to t \bar t$

The $t$- and $u$-channel top exchanges now involve contributions from both $t_L$ without form factor
and $t_R$ with its form factor.
However the standard $Zt_Rt_R$ coupling is already (about 2 times) weaker
than the $Zt_Lt_L$ one and as these couplings appear squared in these amplitudes
(so at 4th power in the cross section),
the suppression of the $t_R$ coupling would produce almost no visible effect
in the cross section.
On another hand the $s$-channel now only involves the Higgs exchange and requires the same study
as in the above $WW$ case.

Considering again the two above extreme cases (suppression of the $t_R$ coupling
with a minimal change of Higgs coupling or the same with, in addition, a suppressed effective top mass)
we would obtain high energy ratio values of the order of 1 or of 0.7, respectively; see left
and right panels od Figure 3 respectively.
However this $ZZ$ channel is not directly observable because one should add the
$\gamma\gamma, \gamma Z$ background processes; these are indicated in the panels of
 Figure 3 as ``$ZZ$+back".

\item
(c) $\gamma\gamma, \gamma Z \to t \bar t$

These subprocesses should be added to the $ZZ$ one when considering
$e^-e^+\to e^-e^+ + t \bar t$. They do not involve tree level Higgs exchange,
so they are less model dependent. Their $t_R$ suppression effect is rather
similar to what we have seen in the above real $\gamma\gamma$ case, leading to a
reduction factor of the order of 0.5. However if we apply the effective top mass
rule, the $Z\gamma$ and $\gamma Z$ channels which have a large $Z_L$ contribution
proportional to the top mass are strongly affected and this leads to a ratio
of the order of 0.3.

\end{itemize}

The addition of these various transverse and longitudinal contributions
to these four processes, with different energy dependencies, finally produces a ratio
which is very sensitive to the details of the non standard effects.
The relative importance of the background with respect to $ZZ$ can be varied by applying
angular cuts on the final $e^-e^+$ pair, but in any case, we find a global
reduction factor respectively of the order of 0.5 for the $t_R$ coupling suppression
and of 0.4 when one adds the effective mass rule as one can see in the right panel Figure 3.

\subsection{OTHER PROCESSES}

We have seen that the main $t\bar t$ production processes
in $e^-e^+$, gluon-gluon, and $\gamma\gamma$ collisions
are clearly and directly sensitive to the $t_R$ form factor suppression.
On another hand, we have seen that the $WW, ~ZZ$ fusion processes are in addition also
sensitive to the H compositeness structure and that this could lead to
further suppression effects. We have considered the possibility
of a (very important) scale dependent top mass $m_t(s)$ which strongly affects
this fusion class of processes.
But it would not affect the
above main $t\bar t$ production processes (in $e^-e^+$, gluon-gluon,
and $\gamma\gamma$ collisions) because at high energy
the top mass has a negligible effect in these processes.

We can also mention that other simple top production mechanisms
like $q\bar q \to \gamma,Z,H \to t \bar t$ are less affected by the $t_R$
form factor suppression at high energy, because of particular quark
couplings combinations; while in $q\bar q'\to W \to t \bar b$ a similar consequence is induced
 by the high energy $t_L$-dominance, with the unmodified left-handed W coupling.

We have nevertheless considered the $bg\to tW^-$ process (see \cite{Rindani}
for recent studies) for which the ($b_-g_+ \to t_RW^-_{long}$) helicity
amplitude, although involving the left-handed $W$ coupling,
is directly proportional to the top mass coming from the u-channel top exchange
(see \cite{bgtw}). This property can be understood by using the equivalence
with the $bg\to tG^-$ and the Goldstone coupling proportional to $m_t$.

Applying the replacement (\ref{mtop-replacement}) of $m_t$ by the effective top mass,
would have the effect of essentially suppressing
this longitudinal W contribution at high energy.
Adding the unaffected transverse W contributions, would then lead to a resulting total
unpolarized $bg\to tW^-$ cross  reduced by a factor 0.7 with respect to the SM case,
in the central region at energies above 1 TeV; see Figure 4. Let us add
that one should not forget that the Goldstone equivalence is only valid in
gauge theories so that it may not apply to any kind of effective Higgs description.

There are however other
more complex processes which are directly sensitive to the top
mass. Such processes are:
\begin{itemize}

\item
\underline{$e^-e^+\to t\bar t H$}.

 This is the simplest one, whose  Born contributions involve  production of
$t\bar t$ through $\gamma,Z$ exchange in the s-channel and
 emission of  H by  $t$,   $\bar t$,  or the intermediate $Z$; see \cite{ttH}.
Note that the first two cases involving the $Ht\bar t$ vertex,
are directly sensitive to the top mass. Here  again we look at the two
possibilities, either the simple suppression of $t_R$ couplings
to $\gamma,~Z$ given by  (\ref{gRVt-s-replacement}, \ref{gRVt-x-replacement})
(keeping unchanged the standard $Ht\bar t$ couplings)
or by imposing in addition the $m_t(s)$ suppression through (\ref{mtop-replacement}).
The results are shown in Figure 5, with the first option giving  a
reduction by a factor of just above 0.4, and  the second one giving  a stronger
regular decrease.

\item
\underline{$e^-e^+\to t\bar t Z$}, studied in \cite{ttZ}.

Here again the Born contribution is built by starting from the diagram
 $t\bar t$ production through $\gamma,Z$ exchange in the s-channel,
 and inserting to it a $Z$-production vertex.
We thus encounter  the competition of two different kinds of diagrams:
first, $Z$ emission from $t,\bar t$ lines (with $\gamma,Z$ exchange in the s-channel) together
with $Z$ exchange in the s-channel with intermediate $ZZH$ coupling followed by $H\to t\bar t$,
which are very sensitive to the top properties in particular to the top mass;
second, $Z$ emission from $e^-,e^+$ lines (with $\gamma,Z$ exchange in the s-channel
which is the only part slightly sensitive to the $t_R$
couplings).
In Figure 6 we  show both the unpolarized $Z$ production and the longitudinal $Z_L$
case with the same two options of $t_R$ and $m_t(s)$ modifications, as above.
Again one sees that $Z_L$ production (from the first set) has a large sensitivity
to the top mass due to its basic cancellation property (in agreement with Goldstone equivalence)
but this effect is hidden at high energy by the terms coming from the second set which
do not have this property.

\item
\underline{$gg,\gamma\gamma$ $\to t\bar t H$ and $t\bar t Z$}.

In order to explore more directly the final state properties
(especially in the $t\bar t Z$ case)
one should use another production process, for example
in gluon-gluon or in $\gamma\gamma$ collisions which do not involve the above
initial $Z$ emission. In these cases the amplitudes
correspond solely to $t,u$ channel top exchange and (in the gluon-gluon case) to $s$ channel
gluon exchange. So we have considered both $t\bar t H$ and $t\bar t Z$ production
in the gluon-gluon and in the $\gamma\gamma$ cases. As above we have computed
the two types of ratios of new cross sections over standard ones, the first one corresponding
to  gluon, $\gamma$ and $Z$ couplings to $t_R$ being affected by
the effective form factor, and the second type involving, in addition, the reduced
effective top mass.

The results are illustrated in Figure 7-10.  For simplicity we have computed the ratios
of differential cross sections computed in the center of the 3-body phase space.
They should be very close to ratios of integrated cross sections with cuts avoiding
collinear domains.
In the $t\bar t H$ case (Figure 7,8) one gets reduction effects similar or even stronger
than the ones obtained in $e^+e^-\to t\bar t H$. In the $t\bar t Z$ case one
gets also similar effects (Figure 9,10) again with the large sensitivity of $Z_L$ production
to the effective top mass. The pure transverse $Z_T$ production is not shown because it is
very little affected and especially not by the reduction of the effective
top mass.

\end{itemize}

\subsection{FINAL REMARK}

One may finally worry about the possibility of partial $t_R$ compositeness.
In practice it means that the effective $\gamma t_Rt_R$ and $Zt_Rt_R$ couplings
should be modified by a factor of the type
\bq
\cos\phi+F(s)\sin\phi
\eq
where $\phi$ is the mixing angle (which is equal to $\pi/2$ in the full
compositeness case) of the elementary top quark with the new sector
and $F(s)$ a form factor similar to the one we used above.

In such a case the high energy limits of the ratios considered
in the above sections will lie between
the values obtained in the above full compositeness case
and the value 1 corresponding no mixing with $\phi=0$.

\section{CONCUSIONS}

In this paper we have shown that the assumption of
$t_R$ compositeness (while keeping $t_L$ elementary)
leading to suppressed $\gamma t_Rt_R$ and $Zt_Rt_R$
form factors at high $q^2$, could immediately be checked by
looking at the size of the cross sections and asymmetries
of several basic production processes. The most spectacular
changes with respect to the standard  case are given by
the following values of the ratios of new quantities (with $t_R$ coupling suppressions)
over standard ones:
a factor 1/2 for the total unpolarized $e^-e^+\to t \bar t$ cross section;
a factor 3 for the longitudinal $e^-e^+$ polarization asymmetry
as well as for its forward-backward asymmetry;
a factor 1/2 for the gluon-gluon as well as for the photon-photon
cross sections and for the neutral fusion process
in $e^-e^+$ scattering.

The charged ($WW$) fusion process is more model
dependent as it is very sensitive to the Higgs exchange contribution.
We have illustrated two extreme cases where the ratio can vary from 1 to 0.2 .
In such a composite top model we noticed the possibility of a (very important)
scale dependent top mass which has strong consequences in this $WW$ fusion
process, in $bg\to tW^-_L$ and in other processes like $t \bar t H$
and $t\bar t Z_L$ production in $e^-e^+$,  photon-photon or gluon-gluon
collisions where specific large reduction factors occur.

The above illustrations only
correspond to arbitrary choices of compositeness effects with top form factors
and a scale dependent effective top mass, but not to precise model predictions.
We have only imposed the constraint that no effect (no anomalous coupling) should
appear at low $q^2$.
In practice, in the case where departures from SM predictions would be observed
at high energy in the considered processes, the following strategy could be applied.
First, one should look precisely at the $q^2$ dependence of these
departures, in order to see how they behave at high energy, in particular
if the ratios tend to constant limiting values.
Such precise observations could characterize the type of compositeness model
responsible for these effects.
One should then analyse the shape of these $q^2$ dependencies at
intermediate energy and modelize them
in terms of new physics constituent parameters,
bound state wave functions, resonances etc.
This would imply important works at future $e^-e^+$, $\gamma\gamma$
and hadronic colliders.

After completion of our work, we were informed that a study of the
 effect of $t_R$ compositeness at Tevatron and LHC through an effective
four quark operator had been done in \cite{Tait}.\\


\clearpage

\begin{figure}[t]
\vspace*{-1.cm}
\[
\epsfig{file=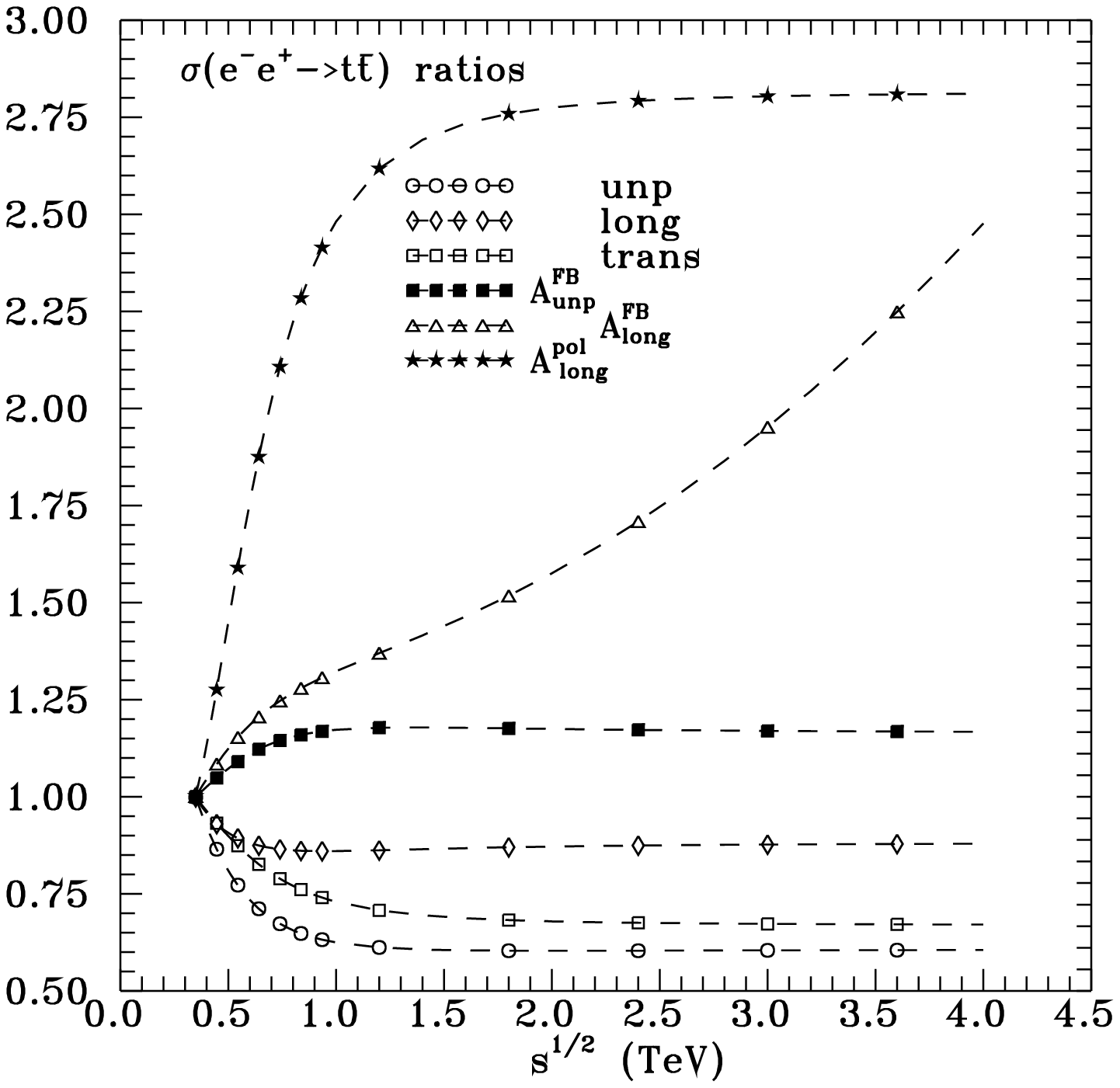, height=8.cm}
\]
\vspace{-1.cm}
\caption[1]{The energy dependencies of the
ratios of the cross sections $\sigma(e^-e^+ \to \gamma,Z \to t \bar t)$
with respect to the SM predictions, involving the $g^R_{Vt}$
form factor effects of (\ref{gRVt-s-replacement}).}
\label{Fg1}
\end{figure}

\begin{figure}[b]
\vspace{-2.cm}
\[
\epsfig{file=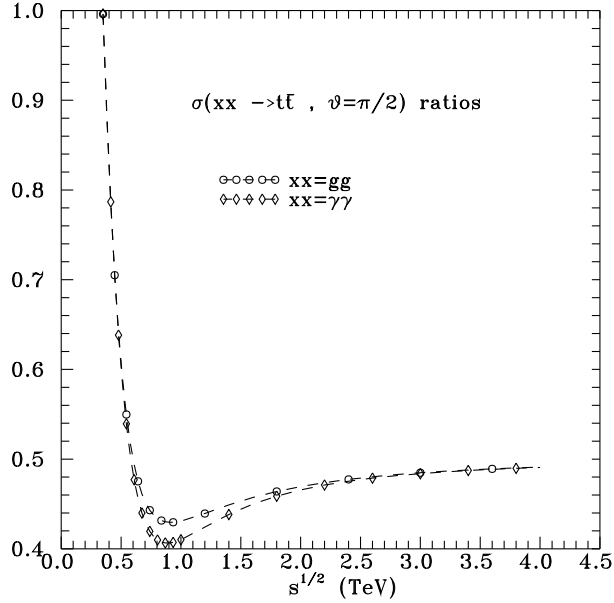, height=8.cm}
\]
\vspace{-0.8cm}
\caption[1]{The energy dependencies of the ratios of the differential cross sections for
$gg \to t \bar t$ and $\gamma\gamma \to t \bar t$ with respect to the SM predictions
at $\theta=\pi/2$, involving the $g^R_{Vt}$ form factor effects
(\ref{gRVt-s-replacement}, \ref{gRVt-x-replacement}). }
\label{Fg2}
\end{figure}

\clearpage

\begin{figure}[t]
\vspace{-1.cm}
\[
\epsfig{file=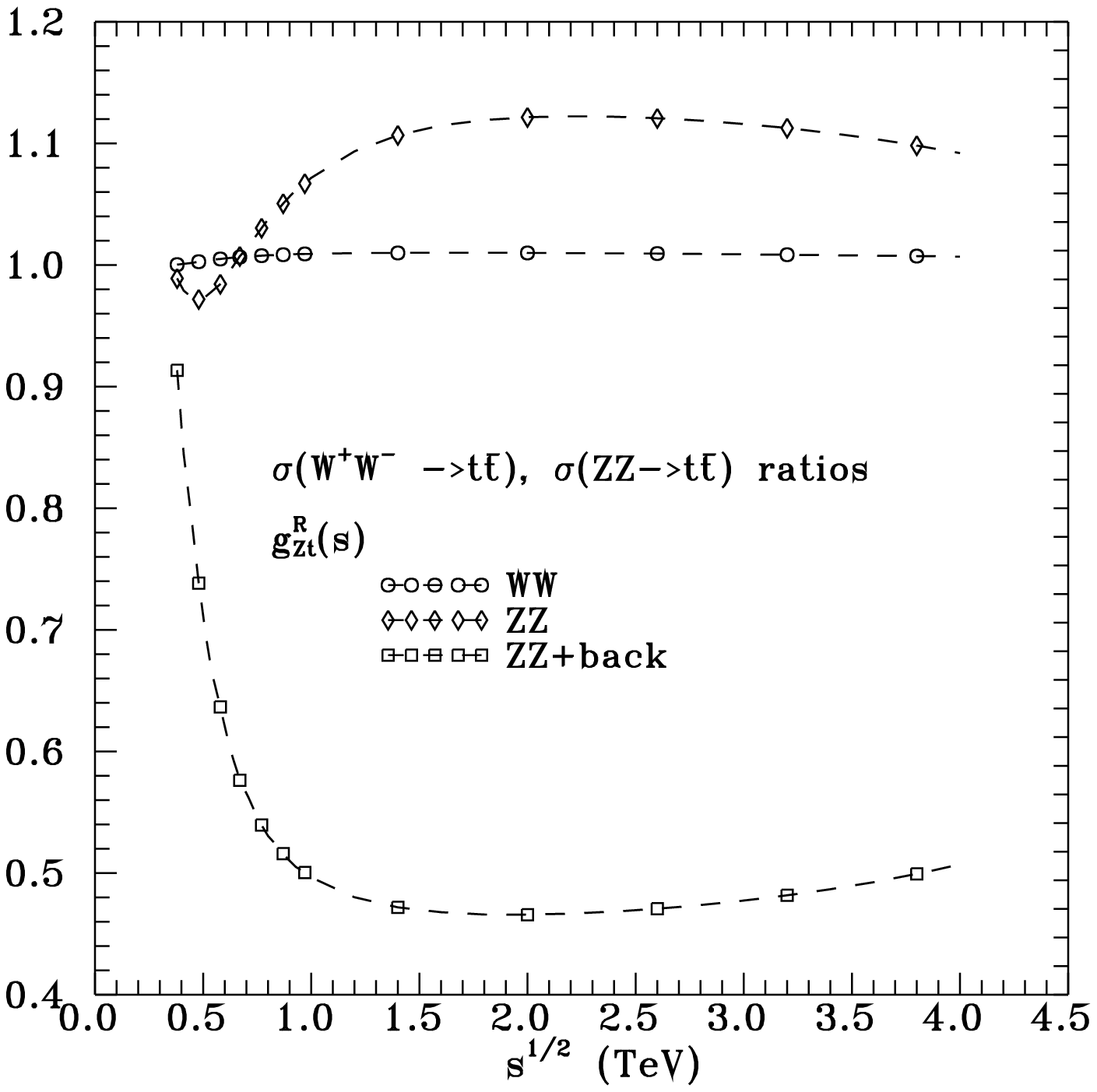, height=7.cm}\hspace{0.5cm}
\epsfig{file=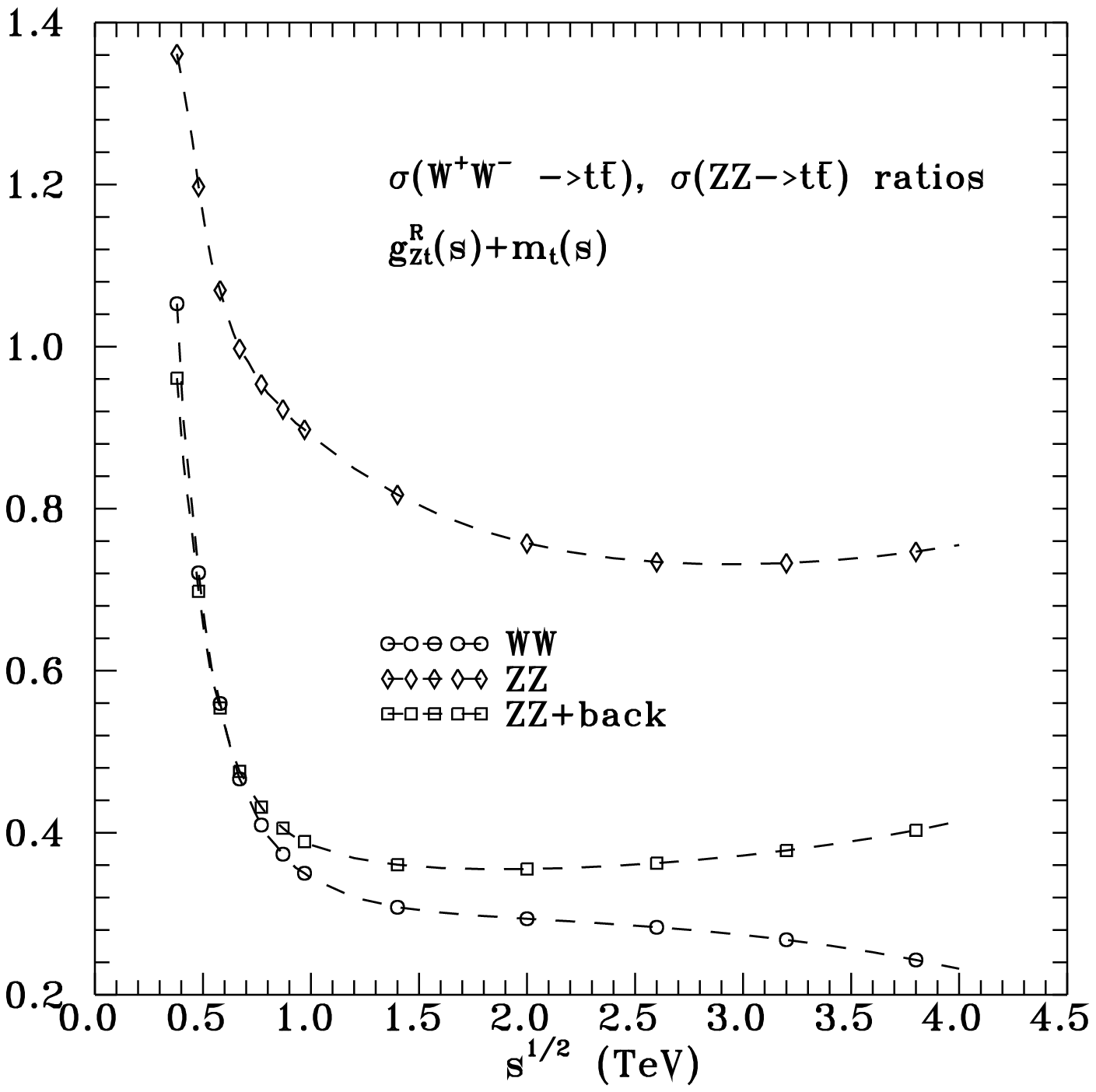,height=7.cm}
\]
\vspace{-0.8cm}
\caption[1]{The energy dependencies of the ratios of $WW$, $ZZ$ and $ZZ$+background
fusion cross sections, with respect
to the SM predictions; see subsection 4.2. Left panel only contains
the $g^R_{Zt}$ effects (\ref{gRVt-s-replacement}, \ref{gRVt-x-replacement}),
while the right panel also includes the $m_t(s)$ effect of (\ref{mtop-replacement}).}
\label{Fg3}
\end{figure}

\begin{figure}[b]
\vspace{-2.cm}
\[
\epsfig{file=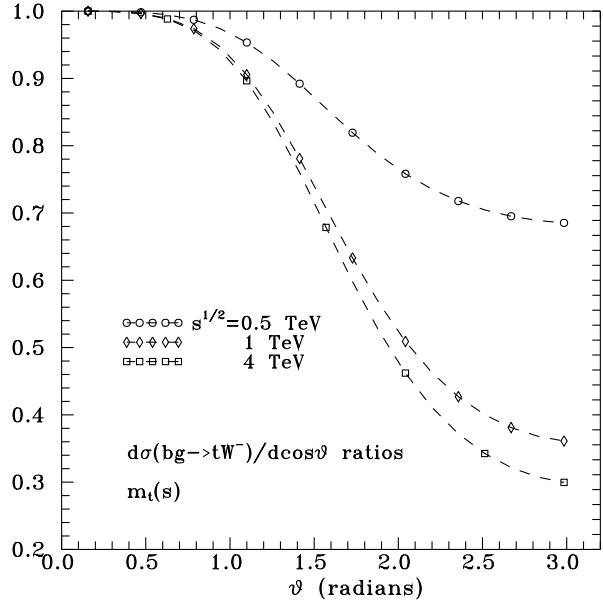, height=8.cm}
\]
\vspace{-0.8cm}
\caption[1]{The angular dependencies of the ratios of the $bg\to tW^-$ differential cross sections,
with respect to the SM predictions.
The form factor effect only depends on $m_t(s)$ of (\ref{mtop-replacement}).}
\label{Fg4}
\end{figure}

\clearpage

\begin{figure}[t]
\vspace*{-1.cm}
\[
\epsfig{file=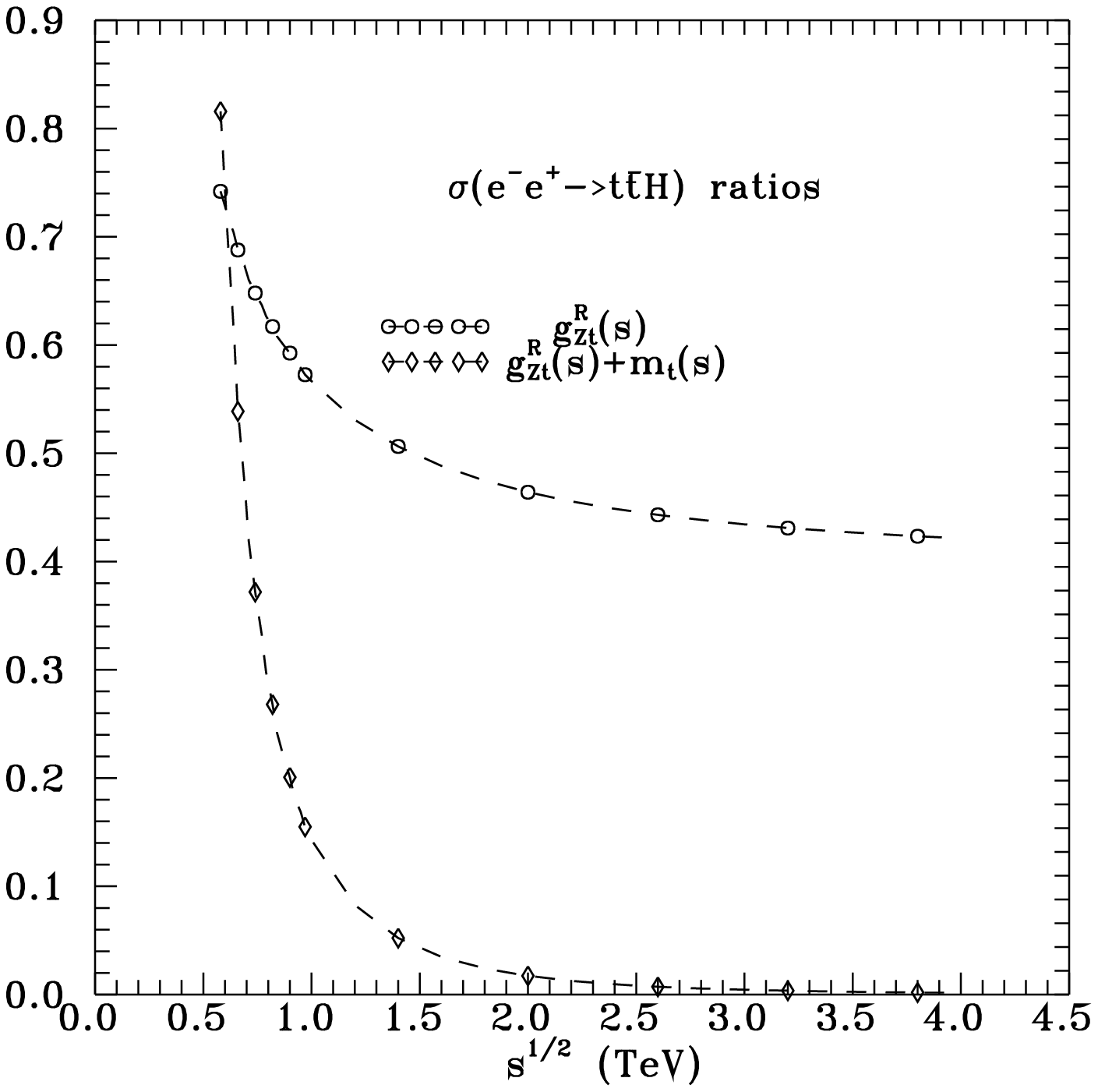, height=8.cm}
\]
\vspace{-1.cm}
\caption[1]{The energy dependencies of the ratios of the cross sections  $e^-e^+\to t\bar t H$,
with respect to the SM predictions. The $g^R_{Zt}$ and the $g^R_{Zt}+m_t(s)$
are obtained from (\ref{gRVt-s-replacement}, \ref{gRVt-x-replacement})
and (\ref{mtop-replacement}).}
\label{Fg5}
\end{figure}

\begin{figure}[b]
\vspace{-2.cm}
\[
\epsfig{file=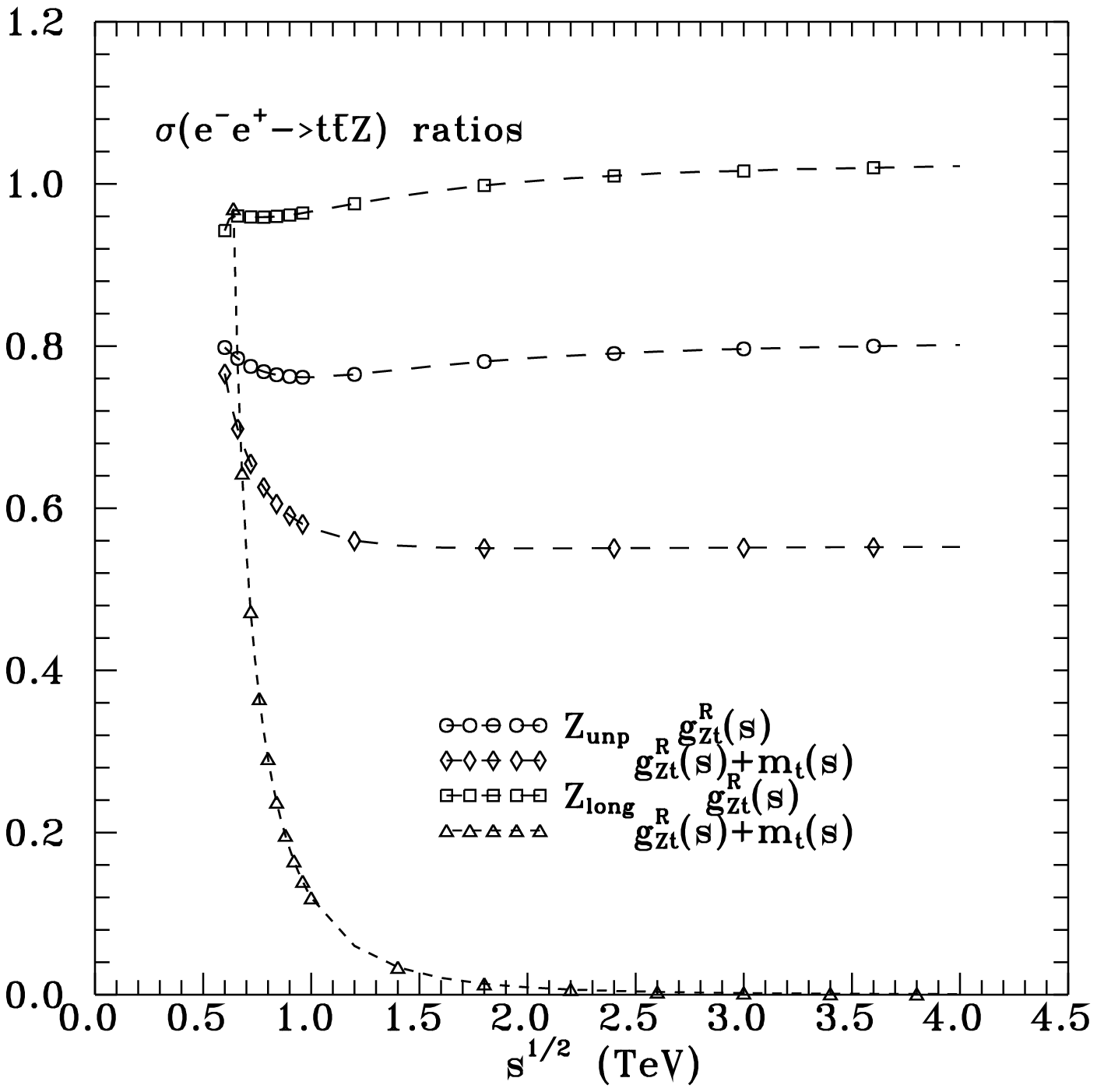, height=8.cm}
\]
\vspace{-0.8cm}
\caption[1]{ The energy dependencies of the ratios of the cross sections $e^-e^+\to t\bar t Z$
for unpolarized $Z$, with respect to the SM predictions. The $g^R_{Zt}$ and the $g^R_{Zt}+m_t(s)$
are obtained from (\ref{gRVt-s-replacement}, \ref{gRVt-x-replacement})
and (\ref{mtop-replacement}). }
\label{Fg6}
\end{figure}

\clearpage

\begin{figure}[t]
\vspace*{-1.cm}
\[
\epsfig{file=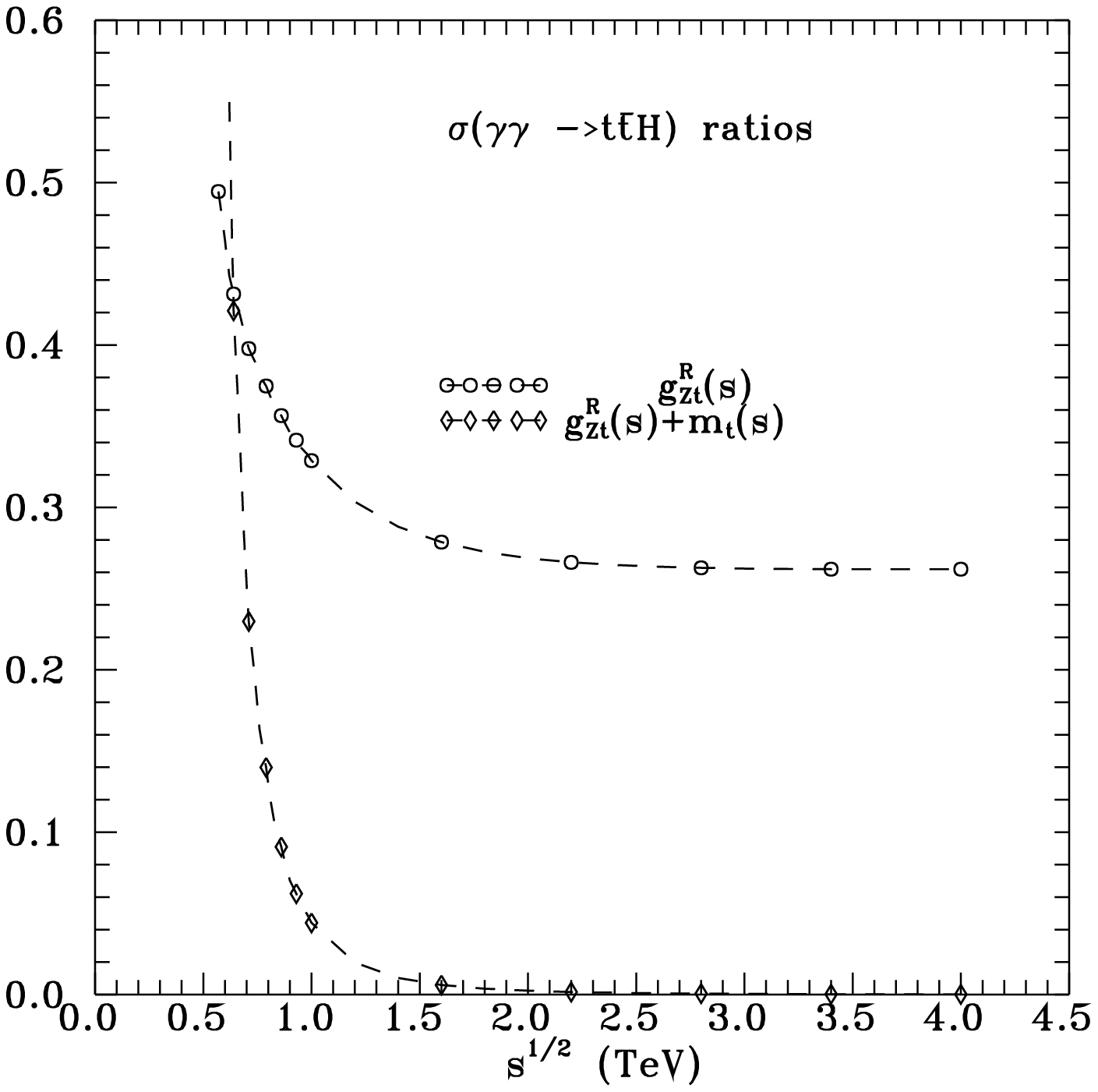, height=8.cm}
\]
\vspace{-1.cm}
\caption[1]{The energy dependencies of the ratios of the cross sections
 $\gamma\gamma \to t \bar t H$ with respect to the SM predictions.
 The $g^R_{Zt}$ and the $g^R_{Zt}+m_t(s)$
are obtained from (\ref{gRVt-s-replacement}, \ref{gRVt-x-replacement})
and (\ref{mtop-replacement}).}
\label{Fg7}
\end{figure}

\begin{figure}[b]
\vspace{-2.cm}
\[
\epsfig{file=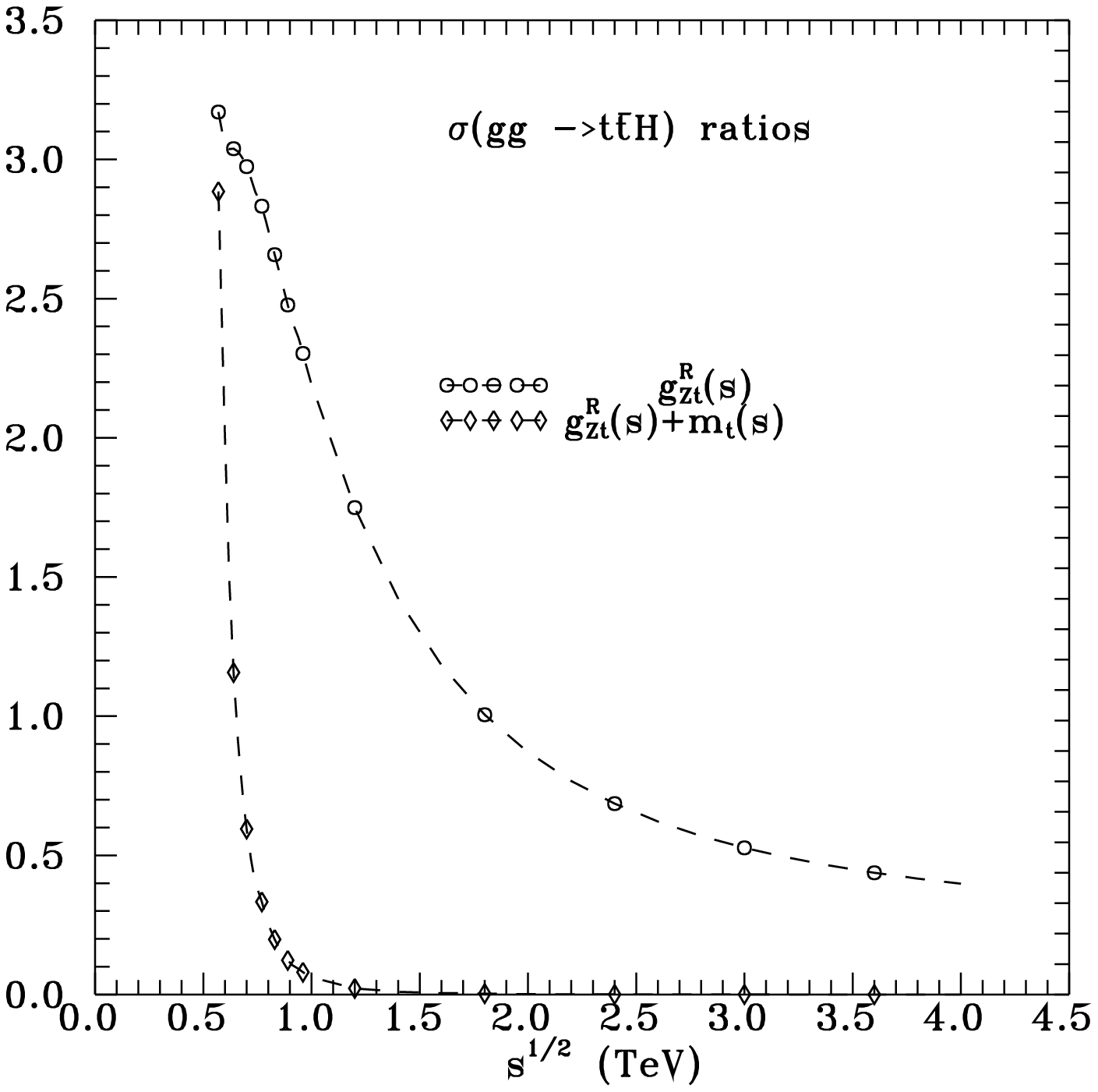, height=8.cm}
\]
\vspace{-0.8cm}
\caption[1]{The energy dependencies of the ratios of the cross sections  $gg \to t \bar t H$
with respect to the SM predictions.  The $g^R_{Zt}$ and the $g^R_{Zt}+m_t(s)$
are obtained from (\ref{gRVt-s-replacement}, \ref{gRVt-x-replacement})
and (\ref{mtop-replacement}). }
\label{Fg8}
\end{figure}

\clearpage

\begin{figure}[t]
\vspace*{-1.cm}
\[
\epsfig{file=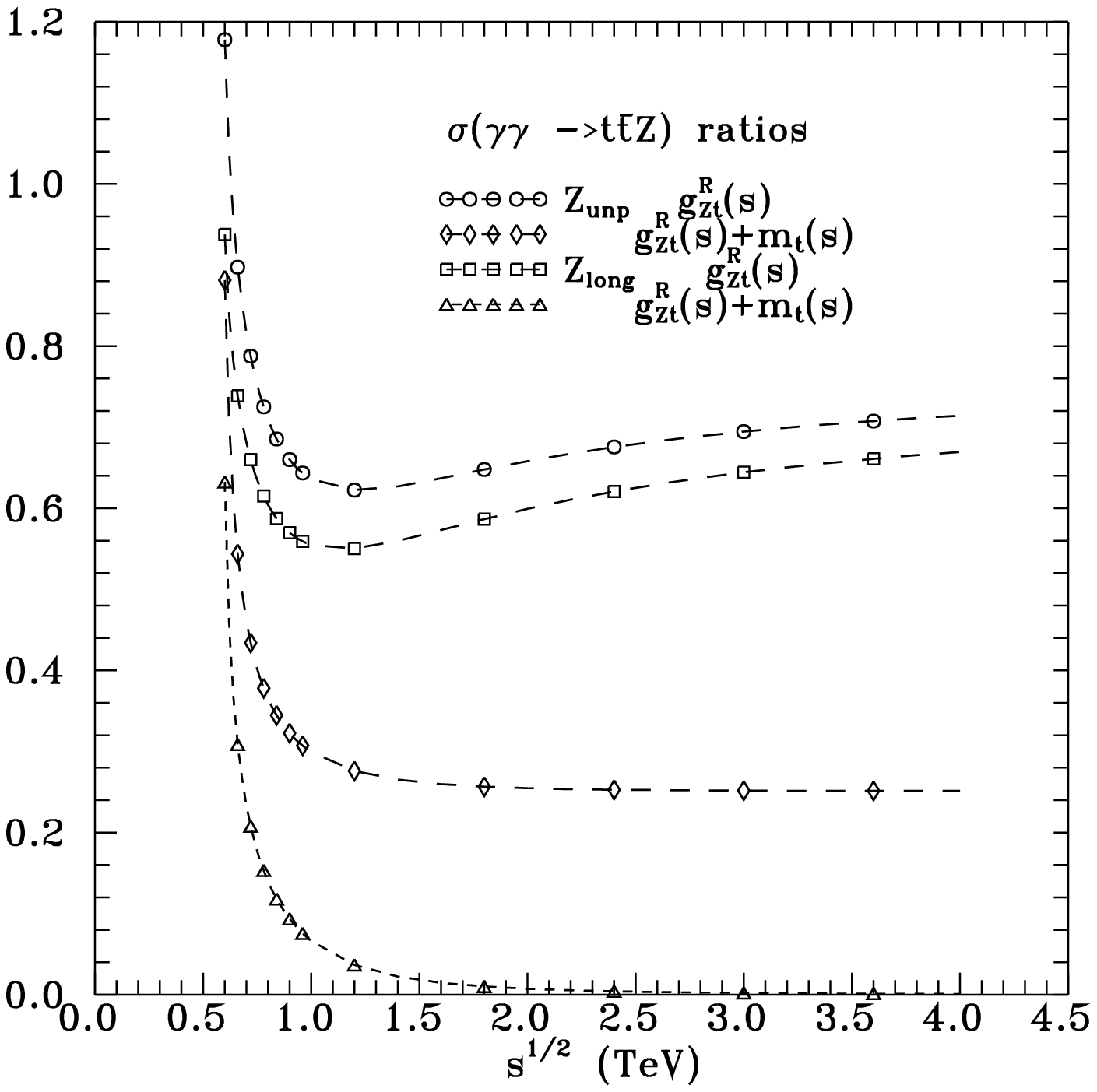, height=8.cm}
\]
\vspace{-1.cm}
\caption[1]{ The energy dependencies of the ratios of the cross sections  $\gamma\gamma \to t \bar t Z$
for unpolarized $Z$, with respect to the SM predictions. The $g^R_{Zt}$ and the $g^R_{Zt}+m_t(s)$
are obtained from (\ref{gRVt-s-replacement}, \ref{gRVt-x-replacement})
and (\ref{mtop-replacement}).}
\label{Fg9}
\end{figure}

\begin{figure}[b]
\vspace{-2.cm}
\[
\epsfig{file=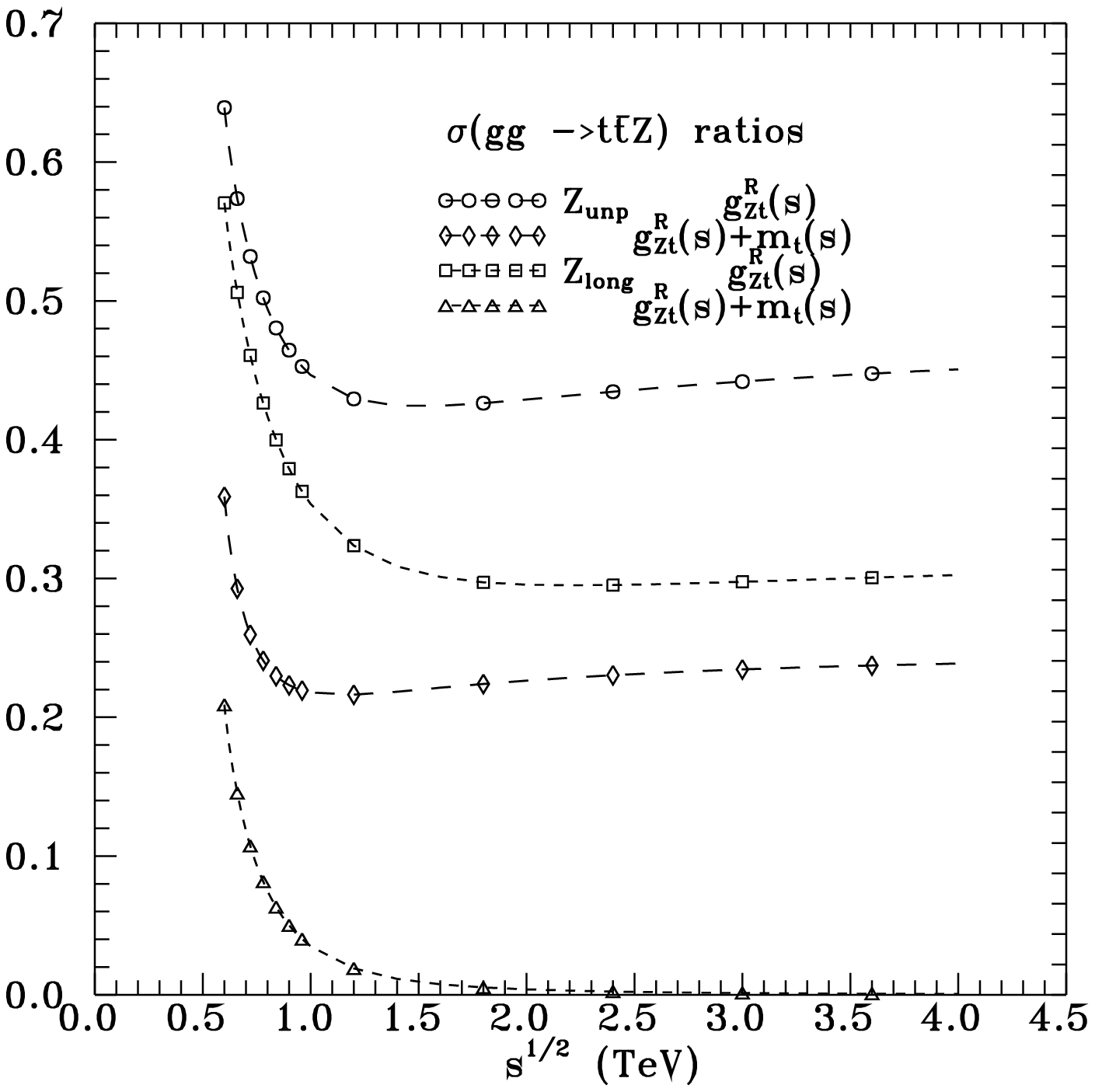, height=8.cm}
\]
\vspace{-0.8cm}
\caption[1]{The energy dependencies of the ratios of the cross sections $gg \to t \bar t Z$
for unpolarized $Z$, with respect to the SM predictions. The $g^R_{Zt}$ and the $g^R_{Zt}+m_t(s)$
are obtained from (\ref{gRVt-s-replacement}, \ref{gRVt-x-replacement})
and (\ref{mtop-replacement}). }
\label{Fg10}
\end{figure}


\begin{thebibliography}{99}
%
%
\bibitem{comp}  H. Terazawa, Y. Chikashige and K. Akama, \pr{D15}{480}{1977};
for other references see
H. Terazawa and M. Yasue, Nonlin.Phenom.Complex Syst. {\bf 19},1(2016);
\jmp{5}{205}{2014}.
%
\bibitem{Hcomp2} D.B. Kaplan and H. Georgi, \pl{136B}{183}{1984}.
%
\bibitem{Hcomp3} K. Agashe, R. Contino and A. Pomarol, \np{B719}{165}{2005}; hep/ph 0412089.
%
\bibitem{partialcomp} R. Contino, T. Kramer, M. Son and R. Sundrum,
J. High Energy Physics {\bf 05}(2007)074.
%
\bibitem{Hcomp4} G. Panico and A. Wulzer, Lect.Notes Phys. {\bf 913},1(2016).
%
\bibitem{Richard} F. Richard, LAL-Orsay-1455, arXiv: 1403.2893 [hep-ph].
%
\bibitem{Fabbrichesi} M. Fabbrichesi, M. Pinamonti
and A. Tonero, \epj{C74}{12,3193}{2014}, arXiv: 1406.5393 [hep-ph].
%
\bibitem{Hff} G.J. Gounaris and F.M. Renard, \pr{D92}{053011}{2015};
 \pr{D93}{093018}{2016},  arXiv: 1601.04142.
%
\bibitem{Htt} G.J. Gounaris and F.M. Renard, \pr{D94}{053009}{2016},
arXiv:1606.08597.
%
\bibitem{bookee}  F.M. Renard, Basics Of Electron Positron Collisions,
Editions Fronti\`eres,1981.
%
\bibitem{Barducci} D. Barducci, S. de Curtis, S; Moretti, G.M. Pruna,
arXiv: 1504.05407 [hep-ph].
%
\bibitem{Kidonakis} N. Kidonakis, arXiv: 1509.07848 [hep-ph].
%
\bibitem{Rindani} S.D. Rindani {\it et al}, \jhep{1510}{180}{2015}; arXiv: 1507.08385 [hep-ph].
%
\bibitem{bgtw} M. Beccaria {\it et al}, \pr{D73}{093001}{2006}; arXiv: 0601175 [hep-ph].
%
\bibitem{Badelek} B. Badelek {\it et al}, \ijmp{A19}{5097}{2004}, arXiv: 0108012[hep-ex].
%
\bibitem{fusion1} M.S.Chanowitz and M.K. Gaillard, \pl{B142}{85}{1984};
G.L. Kane, W.W. Repko and W.B. Rolnick,  \pl{B148}{367}{1984};
S. Dawson, \np{B249}{42}{1985}.
%
\bibitem{fusion2} R.P. Kauffman, \pr{D41}{3343}{1990}.
%
\bibitem{fusion3}  M. Gintner and S. Godfrey, arXiv: 9612342 [hep-ph],
eConf.0960625(1996)STC 130; C.-P. Yuan, \np{B310}{1}{1988}.
%
\bibitem{fusion4}  M. Capdequi-Peyranere {\it et al}, \zp{C41}{1988}{99}.
%
\bibitem{ttH} A. Djouadi, J. Kalinowski and P. Zerwas, \zp{C54}{1992}{255};
 H. Baer, S. Dawson and L. Reina,\pr{D61}{01302}{2000}.
%
%
\bibitem{ttZ} K.Hagiwara, H. Murayama and I. Watanabe, \np{B367}{1991}{267};
U. Baur, A. Juste, L.H. Orr and D. Rainwater,\pr{D71}{054013}{2005}.
%
\bibitem{Tait} Ben Lillie, Jing Shu, Timothy M.P. Tait,  \jhep{0804}{087}{2008}, 
e-Print: arXiv:0712.3057; Kunal Kumar, Tim M.P. Tait, Roberto Vega-Morales
\jhep{0905}{022}{2009}, e-Print: arXiv:0901.3808 

\end{thebibliography}
\end{document}